\documentclass[12pt]{iopart}

\usepackage{iopams}  
\usepackage{epstopdf}
\usepackage{graphicx}
\usepackage{comment}
\usepackage{color}

\begin{document}

\title[Ultrafast scattering dynamics of coherent phonons in Bi$_{1-x}$Sb$_{x}$ in the Weyl ...]{Ultrafast scattering dynamics of coherent phonons in Bi$_{1-x}$Sb$_{x}$ in the Weyl semimetal phase}

\author{Yuta Komori$^1$, Yuta Saito$^2$, Paul Fons$^{2,3}$, Muneaki Hase$^1$}
\address{$^1$Department of Applied Physics, Faculty of Pure and Applied Sciences, University of Tsukuba, 1-1-1 Tennodai, Tsukuba 305-8573, Japan.}
\address{$^2$Device Technology Research Institute, National Institute of Advanced Industrial Science and Technology, Tsukuba Central 5, 1-1-1 Higashi, Tsukuba 305-8565, Japan.}
\address{$^3$Faculty of Science and Technology, Department of Electronics and Electrical Engineering, Keio University,  3-14-1 Hiyoshi, Kohoku-ku, Yokohama, Kanagawa 223-8522, Japan.}
\ead{mhase@bk.tsukuba.ac.jp}

\vspace{10pt}
\begin{indented}
\item[]
\end{indented}

\begin{abstract}
We investigate ultrafast phonon dynamics in the Bi$_{1-x}$Sb$_{x}$ alloy system for various compositions $x$ using a reflective femtosecond pump-probe technique. 
The coherent optical phonons corresponding to the A$_{1g}$ local vibrational modes of Bi-Bi, Bi-Sb, and Sb-Sb are generated and observed in the time domain with a few picoseconds dephasing time. 
The frequencies of the coherent optical phonons were found to change as the Sb composition $x$ was varied, and more importantly, the relaxation time of those phonon modes was dramatically reduced for $x$ values in the range 0.5--0.8. We argue that the phonon relaxation dynamics are not simply governed by alloy scattering, but are significantly modified by anharmonic phonon-phonon scattering with implied minor contributions from electron-phonon scattering in a Weyl-semimetal phase.
\end{abstract}

%
\maketitle
%
%

\section{INTRODUCTION}
In the past decade, topological materials, such as topological insulators (TIs), Dirac semimetals, and Weyl semimetals, have attracted much attention in condensed matter physics owing to growing interest in the fundamental properties of their surface (bulk) band structure \cite{kane2007,hsieh2009observation,nishide2010direct, benia2015surface, zhang2009electronic}, massless quasiparticle dynamics \cite{hsieh2008topological}, spin dynamics \cite{hsieh2009observation,roushan2009topological}, carrier transport \cite{taskin2009quantum}, and many other new physical effects \cite{teo2008surface,zhu2014topological,katayama2018terahertz,li2019terahertz}. 
Topological materials, in particular TIs, have been considered also as promising materials for thermoelectric devices, taking advantage of their intriguing properties, such as low thermal conductivities \cite{samanta2020intrinsically}, and large Seebeck coefficients \cite{Ghaemi2010}.
Bi$_{1-x}$Sb$_{x}$ was first experimentally discovered to be a 3D TI in 2008 (Ref. \cite{hsieh2008topological}), and has been recently predicted to be a Weyl semimetal for some compositions $x$ with a specific atomic arrangement \cite{su2018topological}. In addition, structural changes in Bi$_{1-x}$Sb$_{x}$ depending on the composition $x$ have been theoretically predicted \cite{singh2018elastic,singh2016investigation} and are thought to be an important factor in the stabilization of topological properties in the alloy system.

Although among TIs, Bi$_{2}$Se$_{3}$, Bi$_{2}$Te$_{3}$ and Sb$_{2}$Te$_{3}$ have been extensively investigated from the point of view of ultrafast dynamics 
\cite{Wang2013,Reimann2014,misochko2015polarization,glinka2013ultrafast,sobota2014distinguishing,Mondal2018}, an alloy system like Bi$_{1-x}$Sb$_{x}$ has rarely been studied because of the difficulty in precisely controlling the composition $x$, with which the electronic band structure of the Bi$_{1-x}$Sb$_{x}$ system changes from semiconducting to a simple semimetal (SM) with band crossover \cite{guo2011evolution,nakamura2011topological},
a TI with a gapless surface state, and even to a Weyl semimetal with a Weyl node at a non zone-center position \cite{kim2013dirac, singh2016prediction}.

In practice, the physical information for the Bi$_{1-x}$Sb$_{x}$ system has been limited \cite{hase1998selective,wu2010ultrafast}, and only a few experimental studies 
have been reported on Dirac quasiparticles \cite{hsieh2009observation,nishide2010direct,hsieh2008topological}. 
Moreover, while the properties of lattice vibrations, phonons, have been examined for the Bi$_{1-x}$Sb$_{x}$ system by Raman scattering \cite{zitter1974raman,lannin1979first,lannin1979vibrational}, the presence of instrumental broadening due to slit and laser line widths \cite{lannin1979first} as well as serious background from Rayleigh scattering \cite{lannin1979vibrational} may limit the accuracy of such measurements. These limitations make the indirect measurement of precise phonon lifetimes on picosecond and femtosecond time scales generally difficult to achieve using conventional Raman measurements. Our motivation here is the direct measurement of the dephasing dynamics of coherent phonons on sub-picosecond time scales and this approach offers strong advantages regarding the accuracy of the dephasing time values over those indirectly estimated by line widths from conventional Raman measurements. 

Phonon scattering, such as electron-phonon, anharmonic phonon-phonon, defect-phonon (or alloy) scatterings, are important issues in the field of thermoelectric and topological devices, since phonon scattering determines the thermal conductivity \cite{samanta2020intrinsically} and the mobility of Dirac quasiparticles \cite{hsieh2008topological}. 
Thus, exploring the ultrafast dynamics of, e.g., anharmonic phonon-phonon scattering and alloy scattering, on a sub-picosecond time scale, is a challenging 
issue required for the development of new topological devices.
Such dynamical studies of the phonon scattering processes will offer key physical information not only for device applications, but also for the understanding of the fundamental physical properties of topological alloy systems.
\begin{figure}
   \begin{center}
    \includegraphics[width=8.8cm]{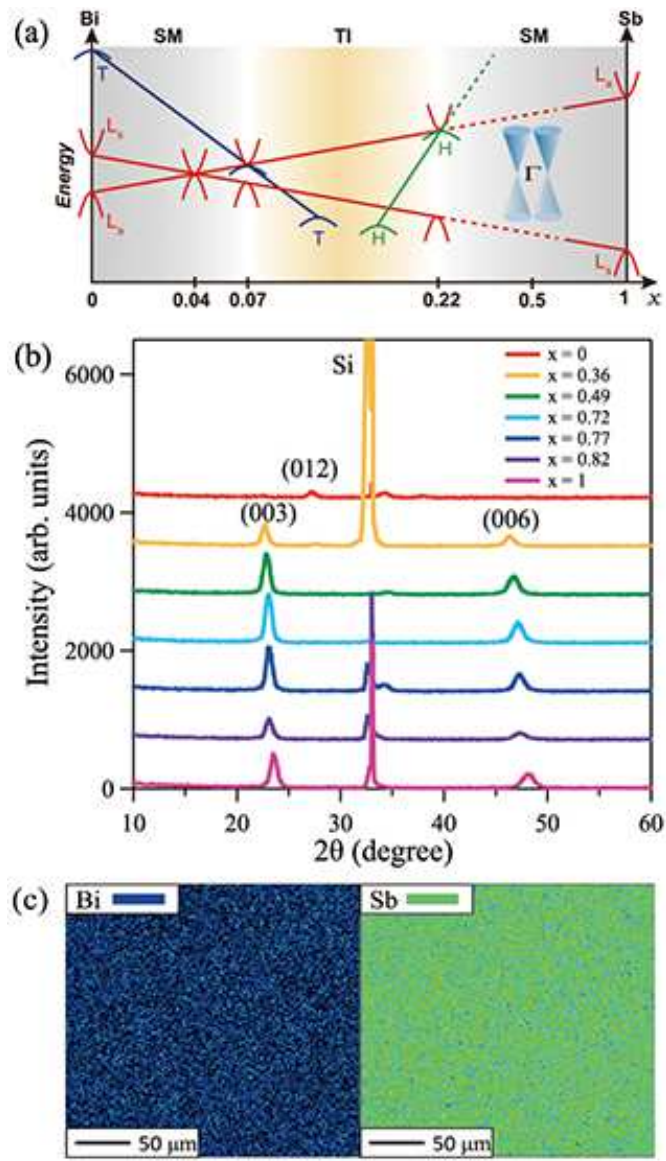}
    \caption{(a) Schematic band structure near the Fermi-level for different values of the Sb concentration $x$, where the alloy changes from a semimetal (SM) to a topological insulator (TI), and then back to a SM phase \cite{kane2007,guo2011evolution}.
    The emergence of the Weyl semimetal phase was theoretically predicted for $x$ = 0.5 and 0.83 near the zone center ($\Gamma$) (Ref. \cite{su2018topological}). (b) XRD patterns obtained from thin film samples with different Sb compositions $x$. (c) 2D mapping (256 $\mu$m $\times$ 256 $\mu$m) of Bi 
    (left) and Sb (right) atoms using EPMA for a Bi$_{0.51}$Sb$_{0.49}$ sample, showing the homogeneous distribution of both elements.
    }
   \label{FIG1}
   \end{center}
  \end{figure}

Here we explore the ultrafast scattering dynamics of coherent phonons in the Bi$_{1-x}$Sb$_{x}$ alloy system for various compositions $x$ by using a femtosecond pump-probe technique with $\approx$20 fs time resolution. 
The coherent optical phonons corresponding to the local vibrations of Bi-Bi, Bi-Sb, and Sb-Sb bonds were detected, 
and the relaxation rate of the Sb-Sb phonon was found to strongly decrease for $x$$\approx$0.5. 
We discuss the phonon relaxation dynamics from the viewpoint of alloy scattering, electron-phonon scattering, and anharmonic phonon-phonon scattering. 

\section{MATERIALS AND METHODS}
\subsection{Samples}
The samples used in this study were highly orientated polycrystalline Bi$_{1-x}$Sb$_{x}$ alloy films prepared by radio frequency magnetron sputtering \cite{saito2020high}. Thin film samples were successfully deposited onto a Si (100) substrate with different compositions $x$ = 0, 0.36, 0.49, 0.72, 0.77, 0.82 and 1.0 by carefully tuning the sputter power of each target.
According to the reported electronic band structures in the literature for the Bi-Sb system, the addition of Sb to Bi shifts the $T$ and upper $L$ bands downwards and the lower $L$ band upwards [figure \ref{FIG1}(a)], resulting in the alloy changing from a semimetal to a topological insulator \cite{guo2011evolution,nakamura2011topological}. 
The topological insulating composition range $x$ is between 0.07 and 0.22, indicating that the $L$-point band is inverted with respect to the electronic band in
Bi \cite{hsieh2008topological,Golin68}. 
For our samples, the $x$ = 0.36 alloy was a semimetal with a narrow band gap (less than several tens of a meV) at the $L$-point \cite{benia2015surface,nakamura2011topological}.
The two samples $x$ = 0.49 and 0.82 are close to the compositions theoretically predicted to form in the Weyl semimetal phase \cite{su2018topological}.
All films were approximately 16-nm-thick and X-ray diffraction (XRD) measurements showed 
that they were highly orientated polycrystalline films as can be seen in figure \ref{FIG1}(b) except for the pure Bi sample which showed the presence of off-orientation peaks. 
The average grain size in the out-of-plane direction estimated from Scherrer's formula \cite{Scherrer1912} was $\sim$ 17 nm, a value
nearly equivalent to the film thickness, suggesting that single grains were formed from the substrate/film interface to the film surface.
This trend was observed regardless of the Sb content, and thus the effect of grain boundary scattering on $x$ is expected to be negligible. It should be noted that the (003) and (006) peaks shift toward higher angle with increasing Sb content suggesting that the lattice constant, $c$, decreases upon alloying with Sb which has a smaller atomic radius.

The composition of the films was evaluated by electron probe micro analysis (EPMA) as shown for the case of $x$ = 0.49 in figure \ref{FIG1}(c). 
To reduce uncertainty in the composition, the value of the Sb content, $x$, was measured at five different positions on the sample, and averaged. We have also measured the transient reflectivity change signal at different sample positions, confirming that the signal did not vary. 
Furthermore, the compositional mapping displayed in figure \ref{FIG1}(c) confirms a uniform distribution of both elements over the substrate surface.
To prevent oxidation, samples were capped by a 20-nm-thick ZnS-SiO$_{2}$ layer without breaking the vacuum. ZnS-SiO$_{2}$ capping layers 
have long been used as a dielectric layer in optical disc applications \cite{Ohshima96} and 
are completely transparent to near infrared light making no contribution to the observed signal. 

\subsection{Experimental methods}
Time-resolved measurements using a reflection-type pump-probe setup were carried out to observe the coherent phonon signal \cite{Zeiger1992,Fukuda}.
The light source was a mode-locked Ti:sapphire laser oscillator with a center wavelength of 830 nm, providing ultrashort pulses 
of $\leq$20 fs duration and operated at a 80 MHz repetition rate. 
The pump and probe beams were focused by an off-axis parabolic mirror on the samples to a diameter of $\approx$19 $\mu$m and $\approx$15 $\mu$m, respectively, assuring nearly homogeneous excitation \cite{Hase13,Minami14,Subkhangulov16}.
The fluence of the pump beam was fixed at $\approx$855 $\mu\mathrm{J/cm}^{2}$, while that of the probe was 
set to $\approx 5\%$ of the pump beam, $\approx$43 $\mu\mathrm{J/cm}^{2}$.
The penetration depth of the laser beam estimated from the absorption coefficients was about 17--20 nm \cite{Hagemann1975}, 
which is larger than the sample thickness ($\approx$16 nm). 
Thus, in the present study the optical excitation was homogeneous over the entire sample thickness and the effects of the penetration depth did not play a role on the 
observed coherent phonon spectra. The delay between the pump and the probe pulses was scanned up to 15 ps by an oscillating 
retroreflector operated at a frequency of 19.5 Hz \cite{hase2012frequency}. The transient reflectivity change ($\Delta R/R$) was recorded as a function 
of pump-probe time delay. 
The time zero was determined as the position of the initial drop, which was confirmed with a Bi semimetal sample \cite{Hase1998}. 
The measurements were performed in air at room temperature.

  \begin{figure*}
    \begin{center}
     \includegraphics[width=16.0cm ]{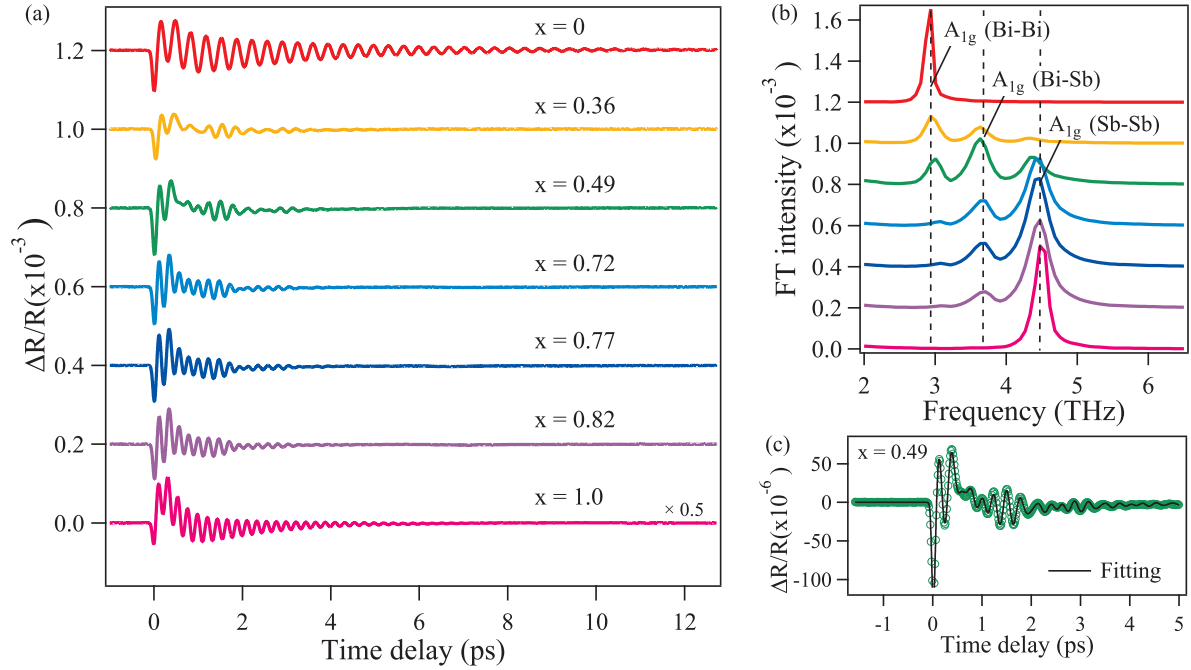}
     \caption{(a)The reflectivity change obtained for Bi$_{1-x}$Sb$_{x}$ alloy films with various compositions at room temperature. 
     (b) Fourier transformed (FT) spectra of time domain data shown in (a). 
     (c) The result of the fit using equation (\ref{EQ1}) for the case of Bi$_{0.51}$Sb$_{0.49}$. 
     }
     \label{FIG2}
    \end{center}
   \end{figure*}
   
\section{RESULTS AND DISCUSSION}
Figure \ref{FIG2}(a) shows the isotropic reflectivity changes ($\Delta R/R$) observed in Bi$_{1-x}$Sb$_{x}$ films with different compositions.
The coherent phonon signal, superimposed on the photoexcited carrier relaxation background, was observed as a function of the time delay.
It is interesting to note that the frequency and oscillation pattern of the coherent phonons differ depending on the composition $x$, under a constant pump fluence. 
We also note that the amplitude of the coherent phonon decreased when the Sb concentration $x$ was in the range of $x$ = 0.36 - 0.82. 
The observed large decrease in amplitude may possibly be due to damping of the coherent lattice vibrations due to atomic disorder, i.e., alloy scattering. 
The relaxation time of the coherent phonons was also found to shorten when the Sb concentration $x$ was in the range of $x$ = 0.36 - 0.82. In particular, 
for $x$ = 0.49, the coherent phonon oscillation was strongly damped for the time delays of $\approx$ 1 and 3 ps (or exhibited a beat pattern), 
implying the existence of multiple phonon modes \cite{hase1998selective}. 

Figure \ref{FIG2}(b) shows the Fourier transformed (FT) spectra obtained from the time domain data shown in figure \ref{FIG2}(a). 
We note here that the non-oscillatory carrier relaxation component was subtracted first and then the FT was carried out to check if the carrier background affected the peak position or peak width of the FT. It was confirmed that the peak position and width did not change with and without subtracting the carrier background. Thus, to retain the original oscillatory information we have displayed the FT without subtraction of the non-oscillatory component.
In total, three distinct peaks can be observed, which are assigned to the A$_{1g}$ modes of Bi-Bi 
(2.93 THz for $x$ = 0), Bi-Sb (3.63 THz for $x$ = 0.36), Sb-Sb (4.54 THz for $x$ = 1), respectively. 
The peak frequencies observed in the FT spectra obtained from the time domain data are reasonably consistent with Raman scattering data \cite{lannin1979first,zitter1974raman,lannin1979vibrational}. 

In order to estimate the relaxation time of the coherent phonon mode for various Sb concentrations $x$, we carried out curve fitting of the time domain data shown in figure \ref{FIG2}(a) using a linear combination of damped harmonic oscillators and an exponentially decaying function.
For the semimetal systems like Bi and Sb, the oscillatory component is well described by a cosine function under the conditions of the displacive excitation of coherent phonons (DECPs) \cite{Zeiger1992}. 
Thus, the fitting function used was \cite{hase1998selective,Hase05}:
\begin{equation}
\label{dampedharmonic}
\frac{\Delta R}{R} = H(t)\Bigl[\sum_{i=1}^{3} A_{i}\exp(-t/\tau_{i})\cos(\omega_{i}t+\phi_{i}) + A_{c}\exp(-t/\tau_{c})\Bigr],
\label{EQ1}
\end{equation}
where $A_{i}$, $\tau_{i}$, $\omega_{i}$, $\phi_{i}$ are the amplitude, relaxation time, frequency, and initial phase of the coherent phonons, respectively. 
The subscripts $i$ indicates the three A$_{1g}$ modes of Bi-Bi ($i$ = 1), Bi-Sb ($i$ = 2), and Sb-Sb ($i$ = 3). 
$H(t)$ is the Heaviside function convoluted with Gaussian to account for the finite time-resolution, while $A_{c}$ and $\tau_{c}$ are the amplitude and relaxation 
time of the carrier background, respectively. As can be seen in figure \ref{FIG2}(c) the quality of the fit was good. 

\begin{figure}[t]
    \begin{center}
     \includegraphics[width=9.5cm]{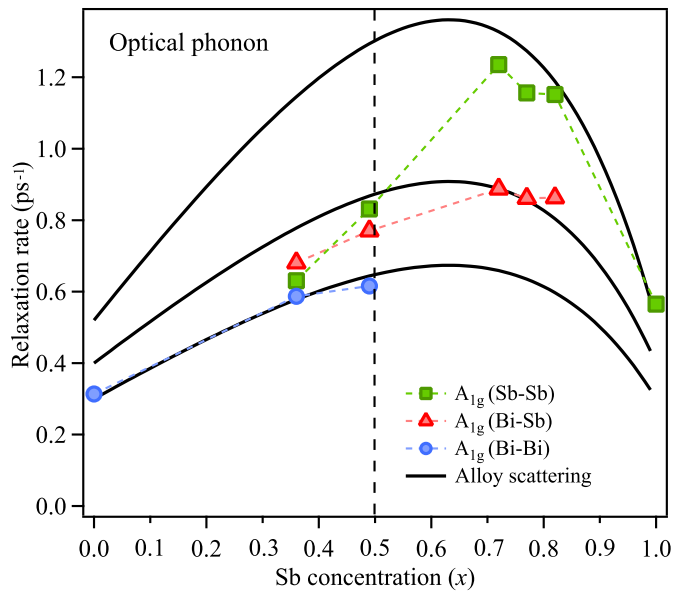}
     \caption{
    The scattering rate of the three different coherent A$_{1g}$ phonons as a function of the Sb concentration $x$. 
    The thick lines indicate the theoretical scattering rate calculated using the alloy scattering model described in the main text, where $\Gamma_{0}=\pi/6$ was fixed for equation (\ref{final}).
    }
     \label{FIG3}
    \end{center}
   \end{figure}

Figure \ref{FIG3} shows the phonon scattering rate $1/\tau$, the inverse of the relaxation time, of the coherent phonon 
oscillations as a function of the Sb concentration ($x$) obtained by a fit using equation (\ref{dampedharmonic}). 
Interestingly, the phonon scattering rate is largest for Sb concentrations of about $x$ = 0.7. 
The scattering rate 1/$\tau$ of the coherent A$_{1g}$ phonon in the Bi$_{1-x}$Sb$_{x}$ alloy is given by the sum of the intrinsic anharmonic 
phonon-phonon scattering $1/\tau_{\rm{Anharmonic}}$, the alloy scattering $1/\tau_{\rm{Alloy}}$, and the electron-phonon scattering $1/\tau_{\rm{e-p}}$. 
The $1/\tau_{\rm{Anharmonic}}$ term is given by the Klemens formula $\gamma_{0}[1 + n(\omega_{\rm{LA, TA}}) + n(\omega_{\rm{A}_{1g}})]$ \cite{Klemens1966}, where $\gamma_{0}$ is the effective anharmonic constant, $n(\omega)$ is the phonon distribution function, and $\omega_{\rm{LA, TA}}$ and $\omega_{\rm{A}_{1g}}$ are the frequencies of the longitudinal acoustic (LA) or transverse acoustic (TA) and the A$_{1g}$ phonons, respectively \cite{Hase1998}. 
Note that the anharmonic phonon--phonon scattering term is assumed to be a constant because the anharmonic decay channel depends mainly on the distribution of lower lying TA and LA phonons and thus on the lattice temperature. 
For the general case with a constant crystal structure, changes in $1/\tau_{\rm{Anharmonic}}$ would be expected to be negligibly small at a constant lattice temperature. 
We have estimated the lattice temperature rise for Bi ($x$ = 0) upon excitation using the two-temperature model (TTM) \cite{Kaganov1957,Allen1987}, which predicts a temperature rise of $\approx$50 K \cite{Mondal2018}. Although the lattice specific heat will slightly increase when the composition ($x$) becomes larger (e.g., by ~20\% as $x$ varies up to 0.12 \cite{Rogacheva2016}), we know from TTM calculations that for larger lattice specific heat values, the lattice temperature rise will decrease. Therefore, we conclude that the lattice temperature is nearly constant or the temperature change is small for the different compositions. 
In the latter sections, however, we will show $1/\tau_{\rm{Anharmonic}}$ varies with the composition $x$. 

\begin{figure}[t]
    \begin{center}
     \includegraphics[width=10.0cm]{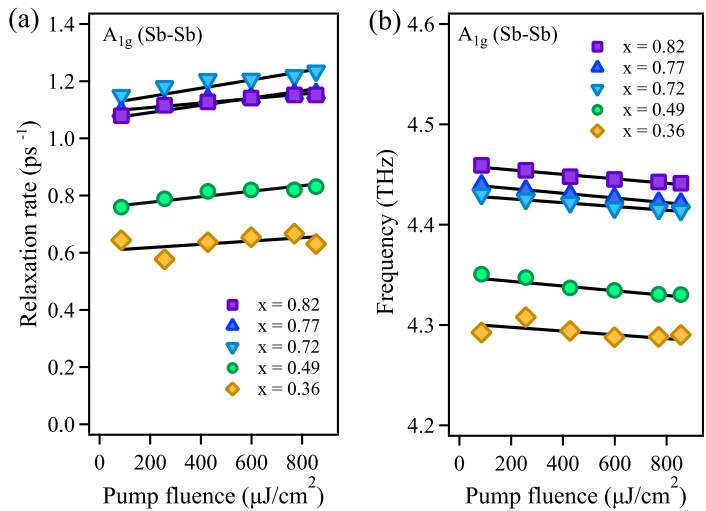}
     \caption{The pump fluence dependence of the relaxation rate (a) and frequency (b) obtained for the A$_{1g}$ phonon 
     mode (Sb-Sb) for various Sb concentration $x$. The solid lines represent a fit with a linear function described in the main text. 
     }
     \label{FIG4}
    \end{center}
   \end{figure}

To examine possible contributions from electron-phonon scattering $1/\tau_{\rm{e-p}}$, the pump fluence dependence of the relaxation rate (the inverse of the dephasing time) and the frequency are presented as a function of the pump fluence $F$ in figure \ref{FIG4}. 
At present, reliable data for the variation of the carrier density over the entire composition $x$ in the Bi$_{1-x}$Sb$_{x}$ system is not available \cite{singh2016prediction}, 
but the photogenerated carrier density is of the same magnitude, as figure \ref{FIG3} was obtained under a constant fluence. 
As electron-phonon scattering would occur near the Fermi-level (E$_{f}$) for the semimetal or Weyl semimetal phase \cite{Parameswaran2014}, we can examine the effect of the carrier density on the electron-phonon scattering near the E$_{f}$ by the analysis shown in figure \ref{FIG4}. 
Here the frequency and relaxation rate represent the real and imaginary parts of the phonon self-energy, respectively, depending on the photo-excited carrier density $n_{c}$ \cite{Cerdeira1973}.
Assuming $n_{c} \propto F$, where $F$ is the pump fluence, we fit the frequency and relaxation rate to $\omega = \omega_{0} + AF$ and $1/\tau = 1/\tau_{0} + BF$, respectively, 
to obtain, for $x$=0.49, $\omega_{0}$ = 4.35 THz, $A$ = -- 0.024 THz$\cdot$cm$^{2}$/mJ and $1/\tau_{0}$  = 0.757 ps$^{-1}$, $B$ = 0.097 ps$^{-1}$cm$^{2}$/mJ, respectively. 
Both the relaxation rate and frequency exhibited a linear and small change (10 \%) as the pump fluence was varied from 86 to 855 $\mu\mathrm{J/cm}^{2}$, as can be seen in the fitting parameters (A and B). The contribution from $1/\tau_{\rm{e-p}}$ is, however, considered to be insufficient to explain the observed larger change (50 \%) in the relaxation rate (figure \ref{FIG3}).
As discussed above, alloy scattering is expected to play a significant role in Bi$_{1-x}$Sb$_{x}$ alloys \cite{Lee2014}, and therefore we examined an alloy scattering model for the phonon scattering rate in figure \ref{FIG3}. 

The scattering rate due to the alloy scattering can be expressed using the phonon frequency $\omega$ \cite{Tamura83,Aksamija2013} as, 
\begin{equation}
\label{taualloy}
\frac{1}{\tau_{\rm{Alloy}}(\omega)}=\frac{\pi}{6}V_{0} \Gamma_{\rm{Alloy}} {\omega}^{2} D(\omega),
\end{equation}
where $V_{0}$ is the volume per atom, $\Gamma_{\rm{Alloy}}$ is an alloy constant, and $D(\omega)$ is the vibrational density of states per unit volume. 
Here, the total vibrational density of states is given by the sum over phonon branch $b$, 
\begin{equation}
\label{DOS}
D(\omega)=\sum_{\mathrm{\it b}}{} \int \frac{d\vec{q}}{{(2\pi)}^{3}}\delta[\omega-\omega_{b}(\vec{q})],
\end{equation}
where $\vec{q}$ is the wavevector. The scattering-rate constant is given by three different contributions, 
\begin{equation}
\label{threecomp}
\Gamma_{\rm{Alloy}}(x)=\Gamma_{\rm{Mass}}(x)+\Gamma_{\rm{Iso}}(x)+\Gamma_{\rm{Strain}}(x),
\end{equation}
where the first contribution is due to the mass difference, the second due the isotope effect, and third strain effects. 
Here we focus on the mass difference, which plays the dominant role in scattering in the alloy Bi$_{1-x}$Sb$_{x}$ \cite{Lee2014}. The mass difference constant is given by,
\begin{equation}
\label{masscon}
\Gamma_{\rm{Mass}}=\sum_{i}f_{i} {\left(1-\frac{M_{i}}{\overline{M}}\right)}^{2},
\end{equation}
where $f_{i}$ is the mass ratio of atoms with mass $M_{i}$, and $\overline{M}$ is the average mass given by $\overline{M}=\sum_{i}f_{i}M_{i}$. The $\Gamma_{\rm{Mass}}(x)$ term depends on the atomic concentration $x$, and therefore is given by 
\begin{eqnarray}
\label{massfanction}
\Gamma_{\rm{Mass}}(x)&=x(1-x)\frac{{(M_{\rm{Sb}}-M_{\rm{Bi}})}^{2}}{{[xM_{\rm{Sb}}+(1-x)M_{\rm{Bi}}]}^{2}} \nonumber \\[0.2cm]
&=x(1-x)\frac{{(121.76u-208.98u)}^{2}}{{[x\cdot121.76u+(1-x)\cdot208.98u]}^{2}} ,
\end{eqnarray}
where the atomic mass of Bi is 208.98$u$, and that of Sb is 121.76$u$, with $u$ being given in unified atomic mass units.
Thus, the $1/ \tau_{\rm{Alloy}}(\omega)$ term shown in equation (\ref{taualloy}) can be written as,
\begin{eqnarray}
\label{kannwa}
\frac{1}{\tau_{\rm{Alloy}}}=
&\frac{\pi}{6}V_{0} {\omega}^{2} D(\omega) \cdot x(1-x) \nonumber \\
& \times \frac{{(121.76u-208.98u)}^{2}}{{[x\cdot121.76u+(1-x)\cdot208.98u]}^{2}}.
\end{eqnarray}
In addition, here, we apply the Einstein model for the $D(\omega)$ term shown in equation (\ref{DOS}) due to the nearly flat dispersion for the optical phonon mode \cite{Kittel}, and thus we assume that all the phonon modes for the optical phonon branch have the same frequency $\omega_{0}$. 
According to the Einstein model, equation (\ref{DOS}) can be reduced to, 
\begin{eqnarray}
\label{dos}
D(\omega)=N \delta(\omega - \omega_{0}) = \rm{const.},
\end{eqnarray}
where $N$ is the number of unit cells. Thus, $D(\omega)$ is independent of the Sb concentration $x$. Based on the above considerations, $1/ \tau_{\rm{Alloy}}$ can be finally written as, 
\begin{eqnarray}
\label{final}
\frac{1}{\tau_{\rm{Alloy}}}=
& \Gamma_{0} {\omega}^{2} \cdot x(1-x) \nonumber \\
& \times \frac{{(121.76u-208.98u)}^{2}}{{[x\cdot121.76u+(1-x)\cdot208.98u]}^{2}},
\end{eqnarray}
where $\Gamma_{0} = \pi V_{0} D(\omega) /6$ is a proportional constant. 
Since the exact values of both $V_{0}$ and $D(\omega)$ constants were not available because of the uncertainly in the crystal structures of the {B}i$_{1-x}${S}b$_{x}$ system, we held $V_{0} D(\omega)$ = 1 for simplicity. In addition, each model curve has a background, assuming $1/\tau_{\rm{Anharmonic}}$ and $1/\tau_{\rm{e-p}}$ do not depend on the composition $x$ as a first trial, i.e., 0.55 ps$^{-1}$ for the Sb-Sb mode,  0.3 ps$^{-1}$ for the Bi-Bi mode, and 0.4 ps$^{-1}$ for the Bi-Sb mode.
The scattering rate thus obtained based on equation (\ref{final}), as a function of the Sb concentration $x$ and the frequency $\omega$ of each optical phonon mode, is plotted in figure \ref{FIG3}. 
As shown in figure \ref{FIG3}, the theoretical scattering rate based on equation (\ref{final}) reproduces the experimental data well, excluding the Sb-Sb vibrational mode around $x=0.5$. 

\begin{figure}
    \begin{center}
     \includegraphics[width=6.0cm]{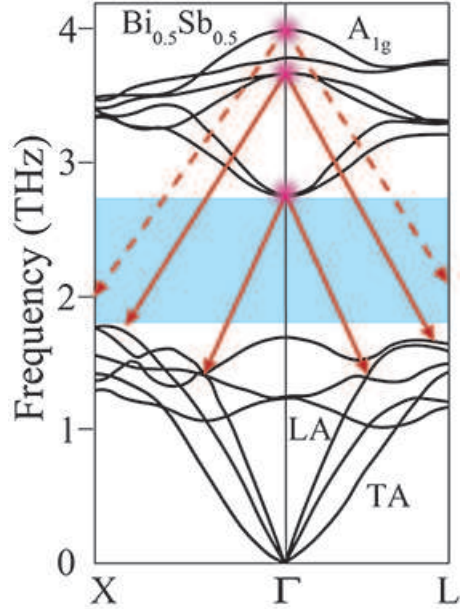}
     \caption{The phonon dispersion in Bi$_{1-x}$Sb$_{x}$ at $x=0.5$. Adapted from Ref. \cite{singh2016investigation}. 
     The solid arrows represent possible anharmonic decay paths from the A$_{1g}$ phonons of the Bi-Bi ($\sim$2.9 THz) and Bi-Sb ($\sim$3.6 THz) 
     local vibrations. The dashed arrows indicate that the anharmonic decay paths from the A$_{1g}$ phonon of the Sb-Sb ($\sim$4.0 THz) are forbidden.
     The rectangle region (light blue) indicates the energy gap between the optical and acoustic phonon branches. }
     \label{FIG5}
    \end{center}
 \end{figure}
We now discuss why the relaxation rate of the Sb-Sb vibrational mode around $x=0.5$ is smaller than the corresponding theoretical value. One plausible explanation is that the anharmonic phonon-phonon scattering rate is accentuated by mass disorder when the phonon frequency decreases, as seen in {S}i$_{1-x}${G}e$_{x}$ alloy systems \cite{Jivtesh2011}. 
In the present study, around $x= 0.5$ where the mass disorder is at a maximum, anharmonic phonon-phonon scattering is accentuated for the lower frequency phonons, specifically the Bi-Bi and Bi-Sb vibrations. 
In this case, the relaxation rate due to the $1/\tau_{\rm{Anharmonic}}$ term should exist for the two modes (Bi-Bi and Bi-Sb vibrations) around $x= 0.5$, while the $1/\tau_{\rm{Anharmonic}}$ term for the Sb-Sb vibrational mode should decrease. 

To test for the above possible scenario based on anharmonic phonon-phonon decay channels in {B}i$_{1-x}${S}b$_{x}$ around $x=0.5$, 
we extracted the phonon dispersion from the literature as schematically shown in figure \ref{FIG5} (Ref. \cite{singh2016investigation}).
Although there are some discrepancies between the frequencies of the optical phonon modes at the $\Gamma$ point in figure \ref{FIG5} and our observations [figure \ref{FIG2}(b)], due to different lattice temperature and a possible mismatch in the lattice structure, there is a large energy gap between the acoustic and optical phonon branches for 1.8 $< \omega <$ 2.8 THz. 
Therefore, the decay channel of the Sb-Sb A$_{1g}$ mode (4.0 THz) is forbidden (red dashed arrows) while the other two optical phonon modes of the Bi-Bi and Bi-Sb A$_{1g}$ modes can decay into the two underlying acoustic phonons under the condition that both the phonon energy and the momentum are conserved (red solid arrows). 
Thus, phonon dispersion considerations can support the idea that the $1/\tau_{\rm{Anharmonic}}$ term significantly decreases for the A$_{1g}$ modes of Sb-Sb around $x= 0.5$. 

\begin{figure}
    \begin{center}
     \includegraphics[width=10.0cm]{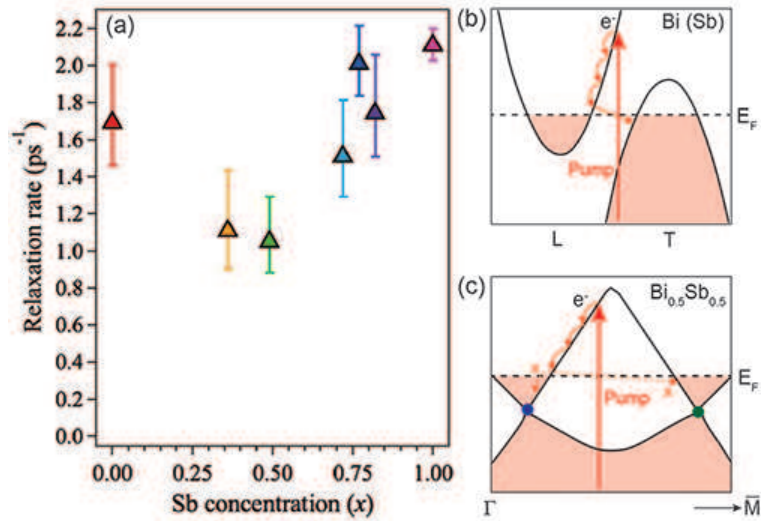}
     \caption{(a) The relaxation rate of the carrier response as a function of the composition. (b) Schematic band structure near the Fermi level for semimetal Bi (Sb), i.e, $x$=0 or 1. The photoexcited electrons (e$^{-}$) relax via electron-phonon intraband scattering, followed by interband recombination. (c) Schematic band structure near the Fermi-level for Weyl semimetal {B}i$_{0.5}${S}b$_{0.5}$ (Adapted from Ref. \cite{su2018topological}). The photoexcited electrons (e$^{-}$) relax via electron-phonon intraband scattering, but interband recombination is prohibited. 
     }
     \label{FIG6}
    \end{center}
 \end{figure}

Another possible explanation is scattering by a quasiparticle coupled to the Sb-Sb bond, which could lead to unique scattering dynamics in the Weyl semimetal phase, such as a decrease in the electron-phonon scattering rate. 
Figure \ref{FIG6} presents the relaxation rate ($1/\tau_{c}$) of the carrier dynamics extracted by the fit in figure \ref{FIG2} as a function of the Sb content $x$. 
The carrier relaxation rate decreases for $x=0.49$, indicating that electron-phonon scattering becomes weak near $x$ = 0.5. 
This result suggests a close relationship between the Weyl semimetal phase ($x$ = 0.5) and the fact that electron-phonon scattering near Weyl points, which are close to the $\Gamma$ point, 
can become weak \cite{Parameswaran2014,Garcia2020} [see also figure \ref{FIG6}(b) and (c)]. 
On the other hand, the contribution from electron-phonon scattering to the phonon dephasing rate is expected to be smaller than that from anharmonic phonon scattering, as discussed in figure \ref{FIG4}, and therefore, the phonon relaxation dynamics are not simply governed by alloy scattering, but are significantly modified by anharmonic phonon-phonon scattering with implied minor contributions from electron-phonon scattering in a Weyl-semimetal phase.
Although exploring such new quasiparticle scattering dynamics will require more experimental and theoretical study, the fact that a reduction in the scattering rate was observed near the Weyl semimetal phase ($x$ = 0.5) suggests a possible contribution from the quasiparticle scattering process described above.

\section{CONCLUSIONS}
In conclusion, we have investigated the dynamics of coherent optical phonons in {B}i$_{1-x}${S}b$_{x}$ for various Sb compositions $x$ by using a femtosecond pump-probe technique. The coherent optical phonons corresponding to the A$_{1g}$ modes of Bi-Bi, Bi-Sb, and Sb-Sb local vibrations were generated and observed in the time domain. 
The frequencies of the coherent optical phonons were found to change with the Sb composition $x$, and more importantly, the relaxation times of these phonons were strongly attenuated for $x$ values in the range 0.5--0.8. We argue that the phonon relaxation dynamics are not simply governed by alloy scattering, e.g., scattering due to mass differences, but are significantly modified by alloy-induced anharmonic phonon-phonon decay channels with minor contributions from electron-phonon scattering in the Weyl semimetal phase, which is expected to appear at $x$ = 0.5.

\ack{
This work was supported by CREST, JST (Grant Number. JPMJCR1875), and JSPS KAKENHI (Grant Numbers. 17H02908 and 19H02619), Japan.
We acknowledge Dr. R. Mondal for helping data analysis and Ms. R. Kondou for sample preparation. }

\section*{References}
\nocite{*}

\end{document}